\documentstyle[12pt,aaspp4,flushrt]{article}

\slugcomment{Submitted to \mnras}

\newcommand{\mbf}[1]{\mbox{\boldmath ${#1}$}}

\newcommand{\ellK}{\mathop{\rm K}\nolimits}

\newcommand{\cE}{{\cal E}}
\newcommand{\Ecrit}{\cE_{\rm crit}}
\newcommand{\rcrit}{r_{\rm crit}}
\newcommand{\mucrit}{\mu_{\rm crit}}
\newcommand{\rt}{r_{\rm t}}
\newcommand{\rp}{r_{\rm p}}
\newcommand{\rh}{r_{\rm h}}
\newcommand{\tp}{t_{\rm p}}
\newcommand{\tr}{t_{\rm r}}
\newcommand{\torb}{t_{\rm orb}}
\newcommand{\ac}{a_{\rm c}}
\newcommand{\amax}{a_{\rm max}}
\newcommand{\amin}{a_{\rm min}}
\newcommand{\Emax}{\cE_{\rm max}}
\newcommand{\Emin}{\cE_{\rm min}}
\newcommand{\Lmax}{L_{\rm max}}
\newcommand{\Lmin}{L_{\rm min}}
\newcommand{\dE}{\Delta \cE}
\newcommand{\dL}{\Delta L}
\newcommand{\dLorb}{\Delta L_{\rm orb}}
\newcommand{\dLprec}{\Delta L_{\rm prec}}
\newcommand{\lnr}{\lambda_{\rm nr}}
\newcommand{\lres}{\lambda_{\rm res}}
\newcommand{\rhos}{\rho_\star}
\newcommand{\Msun}{M_{\sun}}

\newcommand{\psc}{\;{\rm pc}}
\newcommand{\yr}{\;{\rm yr}}

\begin{document}
\null\vskip -0.5truein

\title{Resonant Tidal Disruption in Galactic Nuclei}

\author{\em Kevin P. Rauch\altaffilmark{1} and Brian Ingalls}
\authoremail{rauch@cita.utoronto.ca}

\affil{
Canadian Institute for Theoretical Astrophysics,\\
University of Toronto,\\
60 St. George St., Toronto, On M5S 3H8, Canada.
}
\altaffiltext{1}{Current Address: Dept. of Astronomy, University of Maryland,
College Park, MD 20742-2421.}

\begin{abstract}
It has recently been shown that the rate of angular momentum relaxation
in nearly-Keplerian star clusters is greatly increased by a process termed
resonant relaxation (\cite{raut96}); it was also argued,
via a series of scaling arguments, that tidal disruption of stars in
galactic nuclei containing massive black holes could be noticeably enhanced
by this process. We describe here the results of numerical simulations of
resonant tidal disruption which quantitatively test the predictions made by
Rauch \& Tremaine.
The simulation method is based on an $N$-body routine incorporating cloning of
stars near the loss cone and a semi-relativistic symplectic
integration scheme.  Normalized disruption rates for resonant and
non-resonant nuclei are derived at orbital energies both above
and below the critical energy, and the corresponding angular momentum
distribution functions are found. The black hole mass above which
resonant tidal disruption is quenched by relativistic precession is
determined.  We also briefly describe the discovery of chaos in the 
Wisdom-Holman symplectic integrator applied to highly
eccentric orbits and propose a modified
integration scheme that remains robust under these conditions.

We find that resonant disruption rates exceed their non-resonant counterparts
by an amount consistent with the predictions; in particular, we estimate the
net tidal disruption rate for a fully resonant cluster to be about twice
that of its non-resonant counterpart. No significant
enhancement in rates is observed outside the critical radius. Relativistic 
quenching of the effect
is found to occur for hole masses $M>M_{\rm Q}= (8\pm 3)\times 10^7\Msun$.
The numerical results combined
with the observed properties of galactic nuclei indicate
that for most galaxies 
the resonant enhancement to tidal disruption rates will be very small.
\end{abstract}


\section{Tidal Disruption in Galactic Nuclei}
\label{sec_tdis}

\subsection{Introduction}
\label{sec_intro}

It has long been speculated that massive black holes (MBHs) reside in the
centers of many or most galaxies (e.g., \cite{lyn69})---a notion strongly
supported by recent high-resolution observations of galactic nuclei
(\cite{eckg96}; \cite{vdmet97}; \cite{koret96}; \cite{korr95} and references
therein);
these observations also indicate that the density of stars rises continuously
towards the center of the galaxy. It is nearly certain, therefore, that
tidal disruption of stars by central MBHs actively occurs in many galaxies.

Possible observable consequences of tidal disruption in galactic nuclei have
been proposed by numerous authors. Hills (1975) first suggested that 
the ultimate power source for active galactic nuclei (AGNs)
might be the accretion onto an MBH of gas from tidally disrupted stars
(cf. the related model by \cite{stops92}).
Although subsequent calculations (\cite{fra78}; \cite{mcmlc81}; \cite{duns83})
showed that this mechanism was incapable of
sustaining typical AGN luminosities---the
stellar densities and dispersions required implied
that physical collisions would dominate the mass loss---tidal disruptions may
nonetheless
contribute significantly to the luminosity of some AGNs, particularly
the less energetic variants such as Seyferts and LINERs (low ionization nuclear
emission line regions).
The X-ray outbursts observed in some Seyfert-like nuclei
(\cite{gruet95}; \cite{badkd96}), for example,
could be the result of tidal disruptions. Additional observations 
that can be understood in terms of stellar disruptions include the variations 
in the strength and profile of Balmer lines seen in the Seyfert/LINER NGC 1097 
(\cite{stoet95}) and other active nuclei (\cite{eraet95}),
and the absence of compact nuclear UV sources in $\sim 80\%$
of a sample of LINERs (\cite{eralb95}). The latter, for instance,
was interpreted using a `duty
cycle' model in which the emission line region is powered by episodic
accretion events (produced by tidal disruptions) lasting a few decades each
but active only $\sim 20\%$ of the time on average.
A model by Roos (1992) hypothesizes that broad-line clouds in AGNs are the
remnants of tidal disruptions.  It has also been suggested that
the shell-like structure Sgr A East in the Galactic Center region 
may be the result of a disruption event occurring $\sim 10^4\yr$ ago
(\cite{khom96}).
Other possible signatures of tidal
disruptions include transient flares in otherwise quiescent nuclei
(\cite{ree88}), high-velocity stars produced by tidal dissociation of binaries
(the other star being left bound to the MBH; \cite{hil88}; \cite{hil91}), 
and nuclear enrichment in particular isotopic species 
due to the explosive nucleosynthesis which might occur in relativistic
disruption events (\cite{lumb90}). It has even been suggested (\cite{car92})
that cosmological $\gamma$-ray bursts could be the result of extreme
disruption events in distant nuclei.

Research into tidal disruption can be divided into two classes: studies of the
dynamics of the tidally disrupted debris itself, and studies of the
disruption-induced dynamical evolution of the nuclear star cluster.
The former has been investigated by several groups (e.g., \cite{carl83};
\cite{evak89}; \cite{laget93}; \cite{koc94}), their most basic finding being
that approximately half of the disrupted debris remains bound to the hole
(and is eventually accreted) with the remainder being ejected from the nucleus.
In this paper we undertake an analysis of the latter type: an investigation
of loss cone dynamics under the influence of resonant relaxation
(\cite{raut96}, hereafter RT96), an
enhancement to two-body relaxation that is especially pronounced in
nearly-Keplerian stellar systems.

Unless explicitly overridden, we will employ
geometric units throughout this paper. In these units,
$G=c=M=1$ (where $M$ is the hole mass) and the natural units of time,
distance, and angular momentum are $GM/c^3$, $GM/c^2$, and $GM/c$,
respectively.

\subsection{Review of Non-Resonant Loss Cone Dynamics}
\label{sec_lcone}

The contribution of tidal disruptions driven by two-body relaxation
to the dynamical evolution of star clusters 
has been examined in detail (\cite{frar76};
\cite{ligs77}; \cite{sham78}; \cite{cohk78}). The loss cone itself can be
defined as the region of phase space for which the periapsis of the
stellar orbit, $\rp$, lies inside the star's tidal radius,
$\rt\approx 2 (M/m_\star)^{1/3} R_\star$; these orbits are normally highly
radial. Restated
in terms of angular momentum, loss cone orbits are those with $L<\Lmin$, where
$L=[a(1-e^2)]^{1/2}$
is the angular momentum of a Keplerian orbit with semi-major
axis $a$ and eccentricity $e$, and
$\Lmin=[(2-\rt/a)\,\rt]^{1/2}\approx (2\rt)^{1/2}$ is the angular
momentum of an orbit with periapsis $\rp=\rt$. It is useful to define a
parameter $q=(\dLorb/\Lmin)^2$, where $\dLorb$ is the rms change in 
angular momentum (due to two-body relaxation) per orbital period;
there are then two limiting cases to consider. For $q\ll 1$, stars take many
orbital periods to diffuse across the width of the loss cone and hence stars
entering the loss cone disrupt within an orbital period; this is the ``empty 
loss cone'' or ``diffusion'' limit in which the phase space density drops
exponentially (with scale length $\sim \dLorb$) in the region $L<\Lmin$.
For $q\gg 1$, stars can relax across the loss cone in less than an orbital
period and thus only a fraction $\sim 1/q$ of the stars crossing the loss cone
are disrupted (recall that $L$-space is effectively two dimensional in this
problem); this is the ``full loss cone'' or ``pinhole'' limit for which the
loss cone is dynamically unimportant and phase space densities inside and
outside the loss cone are nearly equal.
Since $\Lmin=\Lmin(\cE)$ and $\dLorb\approx\dLorb(\cE)$
(where $\cE=1/(2a)>0$ is the orbital binding energy)---for
realistic density profiles, the two-body relaxation time 
at fixed $\cE$ depends only weakly on $L$---$q\approx
q(\cE)$ and one can define a ``critical energy,'' $\Ecrit$, 
for which $q(\Ecrit)=1$.  Somewhat more loosely, one can define the boundary
by a critical radius $\rcrit\sim 1/\Ecrit$.

The qualitative influence of the loss cone on the overall structure of the
cluster is now easy to see. 
For $\cE < \Ecrit$ ($r\gtrsim\rcrit$), $q>1$ (the enclosed cluster
mass fraction is larger, orbits are less nearly Keplerian, and hence they
relax relatively faster) and the loss cone has minimal effect on the
distribution function, $f(\cE, L)$.
Outside the critical radius, therefore, the density
profile will follow the classical $\rhos\propto r^{-7/4}$ profile
(\cite{bahw76}; this assumes an isotropic cluster and collisionless cusp
growth---cf. \cite{quihs95}) and the cluster will remain locally isotropic,
i.e., $f(\cE, L)\approx f(\cE)$. Conversely,
for $\cE > \Ecrit$ ($r\lesssim\rcrit$), $q<1$ and the removal of stars by tidal
disruption is significant; the density cusp begins to flatten inside $\rcrit$,
eventually turning over and then
reaching zero at $\rt$ (the detailed calculations
indicate that the flattening inside $\rcrit$ begins logarithmically
slowly---by $r\sim 10\rt$,
the cusp slope flattens by only $\sim 0.2$).
In addition, phase space densities at fixed energy
are severely depleted at low angular
momenta: $f(\cE, L)\approx 0$ for $L<\Lmin$.
For typical ``cuspy'' nuclear
density profiles ($\rhos\propto r^{-\gamma}$ with $1<\gamma<2$,
which encloses the range observed in real galaxies; \cite{lauet95}),
the total rate of disruptions is dominated by the
contribution from stars with energies near $\Ecrit$.
The simulations of Shapiro \& Marchant (1978), for example,
found the mean energy of
disrupted stars to be $\sim 3\,\Ecrit$ (the {\it median} value, on the other
hand, was near $\Ecrit$).

An order of magnitude estimate for $\Ecrit$ follows easily from the definition
$q(\Ecrit)=(\dLorb/\Lmin)^2=1$. First, let $\Lmax=a^{1/2}$ denote the maximum
angular momentum for an orbit with semi-major axis $a$. Then
$\dLorb/\Lmax\sim (\torb/\tr)^{1/2}$ and $\Lmin/\Lmax\sim (2\rt/a)^{1/2}$,
where $\tr$ is the two-body relaxation time (e.g., \cite{spih71}). Assuming in
addition $R_\star\propto m_\star^{2/3}$, $\rhos\propto r^{-7/4}$,
and an orbit-averaged value for $\tr$, we find
\begin{equation}\label{eq_Ecrit}
\Ecrit\sim 2\times 10^{-7}\left(m_\star\over
\Msun\right)^{8/27}M_8^{-1/27}\rho_6^{4/9},
\end{equation}
where $M_8=M/(10^8\Msun)$ and $\rho_6=\rhos(1\psc)/(10^6\Msun\psc^{-3})$.
In physical units, the corresponding value of $\rcrit=1/(2\Ecrit)$ is
\begin{equation}
\rcrit\sim 10\left(m_\star\over
\Msun\right)^{-8/27}M_8^{28/27}\rho_6^{-4/9}\psc.
\end{equation}

We will parameterize the relative importance of tidal disruption in a cluster 
using a dimensionless function $\lambda(\cE)$, defined to be the fraction of
stars at energy $\cE$ consumed in the loss cone each orbital period. Let
$\lambda_0$ denote the theoretical value of $\lambda$ for a fully non-resonant
cluster. In the
pinhole limit ($q\gg 1$), the loss cone is dynamically unimportant and
$\lambda_0$ is simply the fractional area of
the energy hypersurface occupied by the loss cone:
\begin{equation}
\lambda_0\simeq \left(\Lmin\over\Lmax\right)^2,\quad q>1.
\end{equation}
In the empty loss cone limit, $q\ll 1$, the loss rate can be found by solving
the Fokker-Planck equation (e.g., \cite{cohk78}), which yields
\begin{equation}
\lambda_0\simeq \frac{\dLorb^2}{\Lmax^2\ln(\Lmax/\Lmin)},\quad q<1.
\end{equation}
One result of our analysis will be a comparison of $\lambda(\cE)$ between
equivalent resonant and non-resonant systems (\S~\ref{sec_newt}).

\subsection{Resonant Relaxation Summary}
\label{sec_resr}

Under the usual two-body relaxation process, both the energy $\cE$ and angular
momentum $L$ of stellar orbits diffuse (i.e., random walk) slowly through phase
space due to the mutual gravitational 
interactions with other stars in the cluster; 
on average, therefore, the variations in $\cE$ and $L$, $\dE$ and $\dL$,
grow with time $\propto t^{1/2}$, and the energy and angular momentum
relaxation times coincide, $t_\cE\sim t_L$.
As discussed in detail by RT96, resonant relaxation is a
process by which angular momentum relaxation can be dramatically enhanced in
clusters whose mean potential contains `resonant' structure (energy relaxation
is not affected). It operates most
effectively in nearly-Keplerian star clusters, whose
resonant structure consists of spherical symmetry and
equality of the radial and azimuthal orbital frequencies. In this case,
the changes in both
the vector and scalar angular momenta grow approximately linearly with
time---systematically faster than a random walk---and hence the angular
momentum relaxation time is much shorter than the energy relaxation time:
$t_L\ll t_\cE$ (possibly by several orders of magnitude).

Because the nuclear cluster of every galaxy with a central MBH is
nearly-Keplerian close to the black hole, resonant relaxation can have
important consequences for the dynamics of galactic nuclei. Whenever
$t_L < t_H < t_\cE$, for instance (where $t_H$ is the Hubble time),
resonant relaxation will act to isotropize the cluster in less than the age of
the universe, even though the two-body relaxation time is much longer
than $t_H$ (as is typical for galactic nuclei).  In the form of
resonant dynamical friction, resonant relaxation can exert a
controlling influence on the eccentricity evolution
of a massive black hole binary (cf.
\cite{quih97}); it can also rapidly erode the inclination of relatively
massive bodies in nearly-Keplerian disk systems. For details, see RT96.

In this paper we examine in depth another consequence of resonant relaxation:
resonant tidal disruption. In RT96 it was shown that the enhanced relaxation
of $L$ should lead to a large increase of the tidal disruption rate in the
empty loss cone region ($q\ll 1$), by a factor of roughly $q^{-1/2}$;
since disruption rates are not enhanced in the full loss cone region, however,
they estimated that the overall
disruption rate would increase only moderately (a factor of two or three
at most). It was also argued that relativistic precession due to the MBH
should effectively disable resonant tidal disruption for hole masses 
$M\gtrsim 4\times 10^7\Msun$. (Resonant relaxation only operates on
timescales shorter than the orbital precession time, which 
becomes dominated by relativistic effects as the tidal radius approaches the
horizon.) However, resonant relaxation
is a highly non-linear phenomenon, and it is difficult for analytic
calculations to achieve better than order-of-magnitude accuracy;
these numerical estimates are, therefore, rather crude.

In the present work
we discuss the results of detailed numerical simulations designed to
quantitatively test these predictions. 
We calculate both the resonant enhancement to the tidal disruption rate and the
associated change in the cluster distribution function.
The influence of relativistic precession on the results is also investigated,
yielding an improved estimate of the ``quenching mass'' above which resonant
tidal disruption is disabled. We begin
in \S~\ref{sec_method} with a description
of the numerical method. Results are presented in \S~\ref{sec_results},
and concluding discussion is given in \S~\ref{sec_discuss}.
Appendices~\ref{app_grstep} and \ref{app_chaos}
provide additional discussion of technical issues pertaining to
the calculations.

\section{Simulation Method}
\label{sec_method}

\subsection{Numerical Strategy}
\label{sec_nums}

Numerical simulation of resonant tidal disruption is challenging for several
reasons. First is the need to use N-body techniques in a case where the
physical system under study contains many millions of stars. The
non-resonant investigations described in \S~\ref{sec_lcone} avoided this
problem by using a continuum approximation (the Fokker-Planck equation) to
calculate the dynamics; in the resonant case, however, the Fokker-Planck
equation is invalid---it is based on random-walk diffusion of $L$, whereas
resonant relaxation leads to linear growth. Although it should be possible to
modify the Fokker-Planck formalism to accurately account for resonant effects,
this has not yet been accomplished; a particle-based approach is therefore
required. We ease the computational demands by differentiating stars into
two types, test and background (\S~\ref{sec_model}), the former serving as the
tracers of dynamic evolution and the latter representing the mass responsible
for driving that evolution. Further reduction in computational effort was
achieved using a specialized mass spectrum (and associated softening length)
for the background
stars, which allowed fewer stars to be included at large distances
without significantly changing the physical relaxation time (\S~\ref{sec_bg}).

A second constraint is the
high dynamic range needed in $L$-space: determining the resonant modification
to the cluster distribution function, $f(L)$, requires adequately resolving $f$
on scales $\sim\Lmin$; in galactic nuclei, however, $\Lmax/\Lmin\sim 10^5$
for $r\sim\rcrit$. We deal with this problem in two ways: first,
by restricting the simulations to
a manageable subregion of $L$-space defined by $L<L_0$, where $\Lmin\ll L_0
\ll \Lmax$; and second, by {\it cloning} stars approaching $L=\Lmin$
to help reduce statistical fluctuations in this sparsely
populated region of phase space (\S~\ref{sec_tstar}).
Note that attempting to utilize a Fokker-Planck approach would also be
complicated by these low phase space densities present near the loss cone;
previous studies, by contrast, treated the loss cone as a boundary
layer instead of resolving it explicitly.

The radial dynamic range of the test orbits also places stringent demands on
the choice of integrator---another consequence of the extreme mismatch
between $\Lmin$ and $\Lmax$, which implies that loss cone orbits are highly
radial (in our simulations, $e\gtrsim 0.9999$). As discussed in
\S~\ref{sec_tstar}, finding a suitable integration scheme proved an arduous
task; although itself not trouble-free, the Wisdom-Holman symplectic mapping
(\cite{wish91}; this scheme is also known as the mixed-variable symplectic
method) emerged as the clear winner. A slightly modified form of this
method, which we term ``semi-relativistic'' symplectic integration, was used
in simulations examining the influence of general relativity
on resonant tidal disruption.

A cornerstone of our analysis is the use of {\it
comparative} simulations---i.e.,
simulations that were strictly identical apart from
the strength of resonant effects---to determine the importance of resonant
tidal disruption in a particular system.
This `differential measurement' approach removes as much as
possible any systematic biases that may be introduced by the abstractions used
to keep the problem computationally tractable. In any event,
the relatively good agreement
between the theoretical and measured values of quantities such as $\Ecrit$
(\S~\ref{sec_newt}) indicate that such systematic effects are probably small.

\subsection{Model Overview}
\label{sec_model}

The basic computational model contained only two components: a
central black hole and the surrounding nuclear cluster.
Most simulations treated the black hole as a Newtonian point mass; those
including relativistic corrections (\S~\ref{sec_gr}) assumed a Schwarzschild
geometry. In both cases the black hole was held fixed at the center of the
cluster---i.e., we neglect the Brownian motion of the hole, which is dynamically
unimportant in these systems (see \S~\ref{sec_discuss}).
Black hole masses in different simulations
ranged from $10^6$ to $10^8\Msun$ (in the range of those inferred from
observations) and remained fixed for the duration of the run---i.e., we
neglect any evolution of the black hole due to accretion of tidal
debris.

The nuclear cluster was modeled as a halo of bound, $1\,\Msun$ stars with
a distribution function of the
form $f(\cE)\propto \cE^{1/4}$ for $\Emin\le \cE \le \Emax$, and $f(\cE)=0$
elsewhere---in other words, as an isotropic cusp with 
a radial density profile $\rhos\propto r^{-7/4}$ in the range
$\amin \lesssim r \lesssim \amax$,
where $\amin=1/(2 \Emax)$ and $\amax=1/(2\Emin)$ are the
corresponding semi-major axes; this is the Bahcall \& Wolf (1976) solution,
valid for $r\gg \rt$ (see \S~\ref{sec_lcone}).
On physical grounds, $\amin\gtrsim\rt$ and $\amax\lesssim \rh$ are required,
where $\rh$ ($\sim 1/\sigma^2$ for a cluster with velocity dispersion
$\sigma$) is the black hole's dynamical sphere of influence, outside of
which the cusp solution need not apply. In practice, $\amin$ and $\amax$ were
restricted to a narrower band centered on $\rcrit$ in order to minimize
the number of simulated stars; typical values were $\amin\sim \rcrit/30$ and
$\amax\sim 10\,\rcrit$.  The absolute density was chosen to keep the mass
fraction small at the critical radius, a requirement for resonant relaxation
to be effective there; in most cases,
$\rhos(1\psc)\sim 10^3\,\Msun\psc^{-3}$
was used.  Typical galactic nuclei have
$\rhos(1\psc)\sim 10^4-10^6\,\Msun\psc^{-3}$ (cf. \S~\ref{sec_discuss}).

Although the use of N-body techniques in our simulations
was mandatory (\S~\ref{sec_nums}),
a fully self-consistent N-body integration would have been wasteful;
since there is
minimal dynamical evolution of the nucleus as a whole during the course of the
simulations, following the detailed trajectory of every cluster star
is unnecessary. For this reason the simulations separated the cluster into  
two types of stars, test and background, according to the treatment of
their dynamical evolution.

Conceptually the background stars represent the mass of the
cluster, providing the perturbing force felt by the test stars; 
the test stars, by contrast, are the dynamic tracers of the
cluster's equilibrium distribution and tidal disruption rate. 
Hence only the test stars' dynamical evolution is considered---they feel
the forces of the black hole and the background stars (but not of the other
test stars), and their orbits relax nearly as they would in a fully
self-consistent integration; background stars feel only the central force and
execute unperturbed Keplerian motion for the duration of the simulation.
If there are $n$ test stars and $N\gg n$ background stars,
the force calculation 
in this scheme is then $O(nN)$ instead of the much larger $O((n+N)^2)$ for a
self-consistent calculation.  The detailed handling of each is explained in
the following sections.

\subsection{Background Cluster Specifics}
\label{sec_bg}

Given $\amin$, $\amax$, and
$\rho_6\equiv \rhos(1\psc)/(10^6\,\Msun\psc^{-3})$,
the number of background stars, $N$, is known. Even at the rather low
simulated densities, however, there were too many background stars
to include individually---most clusters having $N\sim 10^5-10^6$ and
the practical limit being a few thousand.
The simplest solution to this problem would be to increase the mass of every
background star in proportion to their total overabundance; we adopted a 
more flexible strategy in which mass was made
a function of the star's (constant)
semi-major axis, the most massive ``stars'' occupying the largest,
most distant orbits.
This approach allowed the background cluster to retain both its
fine-grained structure at small radii, where average interstellar distances
are shortest, and its overall mass, which is dominated by the stars at
large radii.
The original mass density profile was preserved by
concomitantly decreasing the stellar {\it number} density at large radii.
More precisely, the background stars were given a mass spectrum of the form
\begin{equation}
{\tilde m}(a) =\left\{ \begin{array}{ll}
  m_\star, & \amin\le a \le \ac; \\
  m_\star(a/\ac)^{5/4}, & \ac \le a \le \amax.
  \end{array} \right.
\end{equation}
In addition, for $a>\ac$ their number density was forced to be $\propto r^{-3}$,
resulting in a mass density $\propto r^{-7/4}$ at all radii, as desired;
for $a<\ac$, stars had their nominal mass and a number density $\propto
r^{-7/4}$.
The initial orbital velocities
of the more massive stars were also given a moderate tangential bias 
to prevent these masses from reaching small radii, where 
they would severely skew the local density; otherwise the orbits were drawn
isotropically.
The value of $\ac$ was determined by the requirement that $\tilde N$ masses,
drawn at random from the above distribution, have (on average) a combined mass 
equal to that of the actual cluster ($=m_\star N$), where $\tilde N$
was the number of background masses to include in the simulation; most runs
used ${\tilde N}=2000$. All simulations used $m_\star=1\,\Msun$.

Another consideration in setting up the background cluster was the amount
by which to soften the point mass potentials of the background stars, which is
necessary with constant timestep integrators (such as the Wisdom-Holman
mapping) to maintain accuracy during close
encounters (see \S~\ref{sec_tstar} below).
Note first that using equal softening lengths for all stars is
impractical---the background masses spanned several
decades in radius, and using the same softening length for each would
produce either negligible softening for the outermost masses or excessive
softening for the innermost.
We therefore decided to use a softening length proportional to
the semi-major axis of the orbit.
The more massive stars, however (those with $a>\ac$),
were treated a bit differently in this
regard. Because they actually represent a group of discrete stars which
have been condensed into a single mass for computational efficiency (a type of
``poor man's tree code''), the softening length for these objects should take
account of their implied
finite size; this in turn depends on their mass and average
density. Based on the specific form of ${\tilde m}(a)$, the softening
length $b$ used for a particular background mass was:
\begin{equation}
b(a) =0.02 a \times \left\{ \begin{array}{ll}
  1, & \amin\le a \le \ac; \\
  5a/(4\ac+a), & \ac\le a \le \amax.
  \end{array} \right.
\end{equation}
Thus $b\approx 0.1a$ for $a\gg \ac$, which provided a good fit to the actual
mean potential of the massive stellar aggregates.
In terms of $b$ the softened potential was $\Phi=-{\tilde m}/(r^2+b^2)^{1/2}$.

Finally, we comment on how the orbital integrations for the background stars
were performed.
As mentioned earlier the background stars follow strictly Keplerian orbits,
and hence their orbital motion can be calculated analytically.
If the integration timestep is small enough, however, direct integration 
will be faster since one analytic Kepler step is an order of
magnitude slower than one direct step.
To maximize efficiency, the code therefore used direct integration whenever
a few such steps would yield sufficient accuracy; otherwise orbits were
advanced analytically.  Direct integrations employed
a second-order leapfrog step which maintained a relative energy accuracy of
$\lesssim 10^{-3}$ at all times.

\subsection{Test Star Integrations}
\label{sec_tstar}

Because of the special demands on the test star integrator (\S~\ref{sec_nums}),
we tested the relative
performance of a wide variety of methods, including both
direct integration schemes (such as leap frog,
adaptive Runge-Kutta, and Bulirsch-Stoer) applied to either the ordinary or
regularized equations of motion and specialized schemes (such as Ecke's method
and the mixed-variable symplectic method) which model the motion
as a perturbed Keplerian ellipse. The latter have the advantage of being
able to use relatively large step sizes for particles undergoing
nearly-Keplerian motion, as is true of the test stars.
Symplectic integrators have the advantage
that they are normally free of long-term growth of energy error, but only if
a constant stepsize is used---hence they also have the disadvantage of being
non-adaptive.
As the extreme orbital eccentricities in this problem demand that the
integrator be (in some sense) adaptive, non-adaptive direct methods such as
leap frog or its higher-order variants were completely unusable. In the end none
of the direct schemes proved to be competitive with the specialized
ones. Of the two methods of the latter type, Encke's method turned out to be
surprisingly inferior to the mixed-variable symplectic (hereafter, MVS)
method, particularly since only the former can utilize an adaptive stepsize.
Encke's method (e.g., \cite{dan92})
works by recasting the equations of motion in terms of
deviations from a fixed Keplerian reference orbit that is periodically reset
(once per orbit, say). In theory,
since these deviations are normally small and
slowly-varying for nearly-Keplerian
motion they can be integrated more easily than the original equations;
in the limit of unperturbed
Keplerian motion the deviations remain precisely zero and the method is exact
for arbitrarily large stepsizes. The MVS method
(\cite{wish91}; cf. \cite{kinyn91})
can be thought of either as a symplectic analog of Encke's method in which the
reference orbit is reset every time step, or (better) as a specialization of
the leap frog step in which the rectilinear motion between the impulsive
velocity changes is replaced by a Keplerian orbital segment; this method is
also exact in the limit of unperturbed Keplerian motion.
Our testing highlighted the fact that even if the absolute
deviations integrated by Encke's method remain small,
their {\it derivatives} can vary sharply---mainly near periapse.
This implies that the deviation equations must be very carefully integrated
when passing periapse, requiring the use of extremely small stepsizes for highly
eccentric orbits; this probably explains the method's poor performance in our
problem. We therefore chose to use the MVS method for our calculations.
Not even this scheme was problem-free, however, as it was found to inject a
(remarkably!) large amount of spurious energy relaxation into the motion.
Appendix~\ref{app_chaos} discusses the origin of this behavior (which appears to
be dynamical chaos in the MVS integrator) as well as the workaround we used to 
mitigate the problem. The final simulations were not adversely affected by
the malady. A fringe benefit of the MVS scheme is the ease with which tidal
disruptions can be detected; since the spatial motion between velocity
perturbations is purely Keplerian, the minimum radius crossed during each step
can be efficiently determined.

As noted previously it was impractical to simulate all of 
$L$-space in our calculations---even resonant relaxation is too slow
for this to be computationally feasible with N-body methods.
We therefore focussed attention on a more manageable region of phase space
around the loss cone, only considering test orbits with angular
momenta $L<L_0$, where $\Lmin \ll L_0 \ll \Lmax$.  Our estimates of the
resonant modification to $f(L)$ is thus confined to $L<L_0$; since resonant
effects are most dramatic near $\Lmin$, however,
this was not a significant restriction (see \S~\ref{sec_results}).
There is, of course, nothing physically special about this boundary, and
test stars inside the simulation boundary will freely relax across it unless
manually restricted.
We implemented this restriction by considering $L=L_0$ to be a {\it
reflecting} boundary---stars crossing outside $L=L_0$ were simply removed
from the simulation and replaced by {\it clones} (explained below)
crossing the boundary towards the inside. 
In most simulations, $L_0\gtrsim 10\,\Lmin$ was attainable.

To adequately resolve $f(L)$ across the entire $L<L_0$ subspace, the region
was divided into a number of non-overlapping zones
spaced logarithmically in $L$ (typically $\sim 20$),
each covering a factor of $\sim\sqrt{2}$ in $L$.
Determining $f(L)$ (for a fixed energy $\cE$) was done by integrating
a population of test stars, each with initial orbital energy $\cE$, until
they achieved dynamic equilibrium; $f(L)$ was then determined by
computing the average occupancy of each zone (weighted by their relative areas).

The small relative areas enclosed by the innermost zones
makes them subject to large statistical uncertainties (i.e., Poisson
noise) due to the very small average occupancies within these zones.
To achieve comparable
relative errors in all zones, we introduced {\it cloning} of test
stars into the simulations, which attempts to equalize the absolute
occupancies between zones by duplicating stars entering the innermost zones.
Conceptually our cloning procedure is very similar to that used by Shapiro \&
Marchant (1978) in their Monte Carlo loss cone simulations; it differs in that
whereas their cloning occurred in energy space (allowing $f(\cE)$ to be
determined well above the critical energy, $\Ecrit$), ours is done in 
$L$-space.  The cloning process proceeded as follows. 
At the start of a simulation a number of zones were designated
`$n$-tuple cloning zones,'
meaning that any test star relaxing into such a zone from above (i.e., from an
adjacent orbit with larger $L$) was replicated $n$ times; subsequently, the
$n$ clones were integrated precisely like non-clones, except that any clone
passing outside the zone in which it was created
was removed from the simulation. Clones were, however, free to relax
inwards (i.e., towards more radial orbits with smaller $L$)---possibly
creating clones of their own, and so on; those
suffering tidal disruption were also removed without being replaced.
Disrupted {\it non-clones}, on the other hand, were replaced by new test stars 
lying just inside the reflecting boundary.
Note that although the number of clones is dynamic,
the number of non-clones (and thus the average test star ``potential'')
remains fixed. The system is therefore dynamically closed and
the zonal occupations come into statistical equilibrium 
after a few crossing times. 
(Recall that since test stars do not feel the gravitational forces of the
other test stars, cloning introduces no dynamical feedback into the system.)
We found that great care was needed in creating the clones
to avoid biasing the calculated distribution function. In particular,
by construction all stars undergoing cloning have $dL/dt<0$ (their
orbits became more radial during the preceding integration step), and not only
must the newly-created clones share this property, the distribution of their
$dL/dt$ values must match that of the uncloned stars. In practice this was
achieved by creating candidate clones, integrating them {\it backwards} one
timestep, and rejecting any that did not cross the cloning boundary during
this step (as the original star had); otherwise the only physical variables
the clones had in common with their parents were
their energies $\cE$ and (scalar) angular momenta $L$.

We tested the reflection and cloning schemes by performing a preliminary
simulation in which tidal disruption was artificially turned off (by simply
not checking for it). Since the perturbing potential provided by the
background cluster is nearly isotropic, in equilibrium the test stars should
exhibit a flat distribution function, $f(L)=const.$ The calculated $f(L)$ was
constant to within $\approx 3\%$, which is less than
the statistical uncertainty of the results presented in \S~\ref{sec_results}.
Other checks---e.g., that the mean density profile of the background cluster
was indeed $\propto r^{-7/4}$---were also done and all returned nominal results.

\section{Results}
\label{sec_results}

The consequences of resonant tidal disruption were examined through a series
of comparative resonant and non-resonant simulations that were otherwise
identical.  The non-resonant results corresponding to a
particular resonant simulation were created by redoing the resonant run while
artificially adding orbital precession (amounting to $\sim 1$ rad per orbit)
to the test stars' motion
to force the precession time down to only a few orbital periods, which is
short enough to extinguish resonant effects; operationally this was done
by adding a small amount of precession each integration step.
This technique allowed a ``differential measurement'' of the disruption rates
and equilibrium distributions to be made, which is
more robust and reliable than attempting to calibrate and match absolute
results obtained from only qualitatively-similar runs.

\subsection{Newtonian Simulations}
\label{sec_newt}

This set of simulations was designed to explore the gross differences in
disruption rates and dynamical equilibrium between resonant and non-resonant
clusters. The runs used a hole mass $M=10^6\Msun$, a value appropriate for
small galactic nuclei such as the Galactic Center or M32 (cf.
Table~\ref{tab_lcprop}). In this case, $\rt\sim 100$
is well outside the horizon and relativistic effects are unimportant; hence
these simulations were performed within an entirely Newtonian framework using
the standard MVS algorithm.

The main simulations consisted of four sets of comparative runs, each at a
fixed orbital energy; three of the four were at an energy above the critical
energy, where resonant enhancements should be prominent. Phase space
parameters for these calculations are shown in Table~\ref{tab_lamb}. The table
lists the orbital energy of the test stars, $\cE$, the maximum
angular momentum followed in the simulation, $L_0$ (i.e., the location of the
reflecting boundary---see \S~\ref{sec_tstar}), the maximum possible angular
momentum at the given energy, $\Lmax$, the average amount of momentum
relaxation suffered per orbit, $\dLorb$, and the typical relaxation
suffered by resonantly relaxed stars over a full precession time, $\dLprec$;
subscripts ``res'' and ``nr'' refer to the values for resonant and non-resonant
runs, respectively. Also given are the results for the normalized
tidal disruption rates,
$\lambda$, with estimated $1-\sigma$ errors given in parentheses ($\lambda_0$
is the `baseline' expectation; see \S~\ref{sec_lcone}). The background cluster
in these simulations had parameters (\S~\ref{sec_bg}) ${\tilde N}=2000$,
$\rho_6=10^{-3}$ ($\rhos\propto r^{-7/4}$),
$\Emin=2\times 10^{-9}$, and $\Emax=5\times 10^{-7}$; it was also found
empirically to have $\Ecrit\approx 2\times 10^{-8}$, in good agreement with
the estimate given in eq.~(\ref{eq_Ecrit}).

The strong resonant enhancement to the local
disruption rate is clearly evident in Table~\ref{tab_lamb}. To within the
accuracy of the results ($\sim 25\%$ at $1-\sigma$),
the ratio of the resonant to non-resonant tidal
disruption rate is given simply by
\begin{equation}
\frac{\lres}{\lnr}= \left\{ \begin{array}{ll}
  \Lmin/\dLorb, & \dLorb\le \Lmin; \\
  1, & \dLorb \ge \Lmin.
  \end{array} \right.
\end{equation}
This agrees qualitatively with the estimate given by RT96, and quantitatively
restricts the previously undetermined numerical coefficient to be very close
to unity.  Combining this result with the analytic estimate for $\lnr$
(\S~\ref{sec_lcone}) implies that the overall rate of tidal disruptions
in a resonant nucleus will be at most three times that for an equivalent
non-resonant nucleus (note: for $\rhos\propto r^{-7/4}$, it is easily shown that
$\dLorb$ is roughly proportional to $\cE^{-1}$), and probably only a factor of
two greater. The precise value depends on the density profile,
the strength of resonant 
effects near $\Ecrit$, the inner radius at which disruptions significantly
deplete the cusp, and the radius at which relativistic effects become
relevant.

\begin{figure}
\plotone{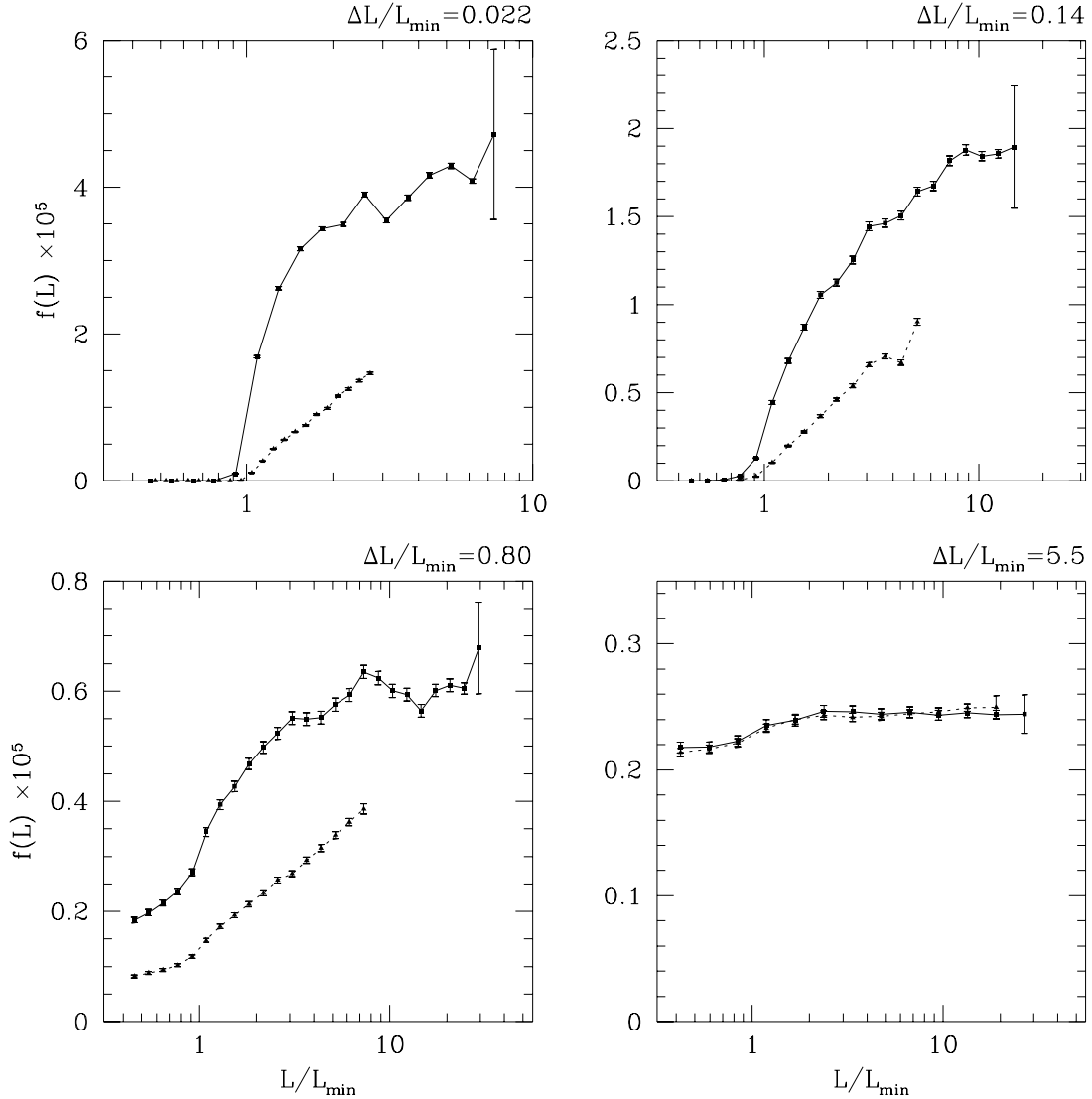}

\caption{
A comparison at several energies of the normalized $L$-space distribution
functions, including approximate 1-$\sigma$ error bars, for equivalent
resonant (solid curves with square points) and non-resonant (dotted curves
with triangular points) clusters.  The large, outermost error bars include
the uncertainty in overall normalization; the remainder show only the formal
statistical uncertainties in each point.
In the empty loss cone limit, $\dLorb/\Lmin\ll 1$, resonant phase space
densities are much higher near the loss cone, implying a large increase in the
local tidal disruption rate relative to non-resonant clusters
(cf. Table~\ref{tab_lamb}).
\label{fig_fLgrid}}
\end{figure}

A comparison of the equilibrium angular momentum distribution functions is shown
in Figure~\ref{fig_fLgrid}, each panel corresponding to a specific value of
$\cE$ (or $\dLorb$, marked); all curves have been normalized to have
unit area in the range $[0, \Lmax]$.
For each curve, the errorbar for the
outermost point includes the overall uncertainty in normalization (resulting
from the need to extrapolate the curves from $L_0$ to $\Lmax$) as well as the
statistical uncertainty in the average bin occupancy, whereas the remaining
points include only the latter. Uncertainty in the normalization of the
resonant curves was
estimated by splitting the difference between two extreme assumptions for how
to extrapolate $f(L)$, namely: (1) $f(L)=const.=f(L_0)$, or (2)
$df/d(\ln L)=const.=df/d(\ln L)_{\rm nr}$ for $L>L_0$.
Note that in agreement with previous studies, the non-resonant $f(L)$ are
well fit by a straight line in semi-log coordinates (outside the loss cone);
for these curves the
extrapolation error was taken to be the formal uncertainty
in the corresponding linear least-squares fit.
The strong resonant increase in phase space density near the loss cone
(which disappears outside the critical radius) is again clearly visible.

\begin{figure}
\plotone{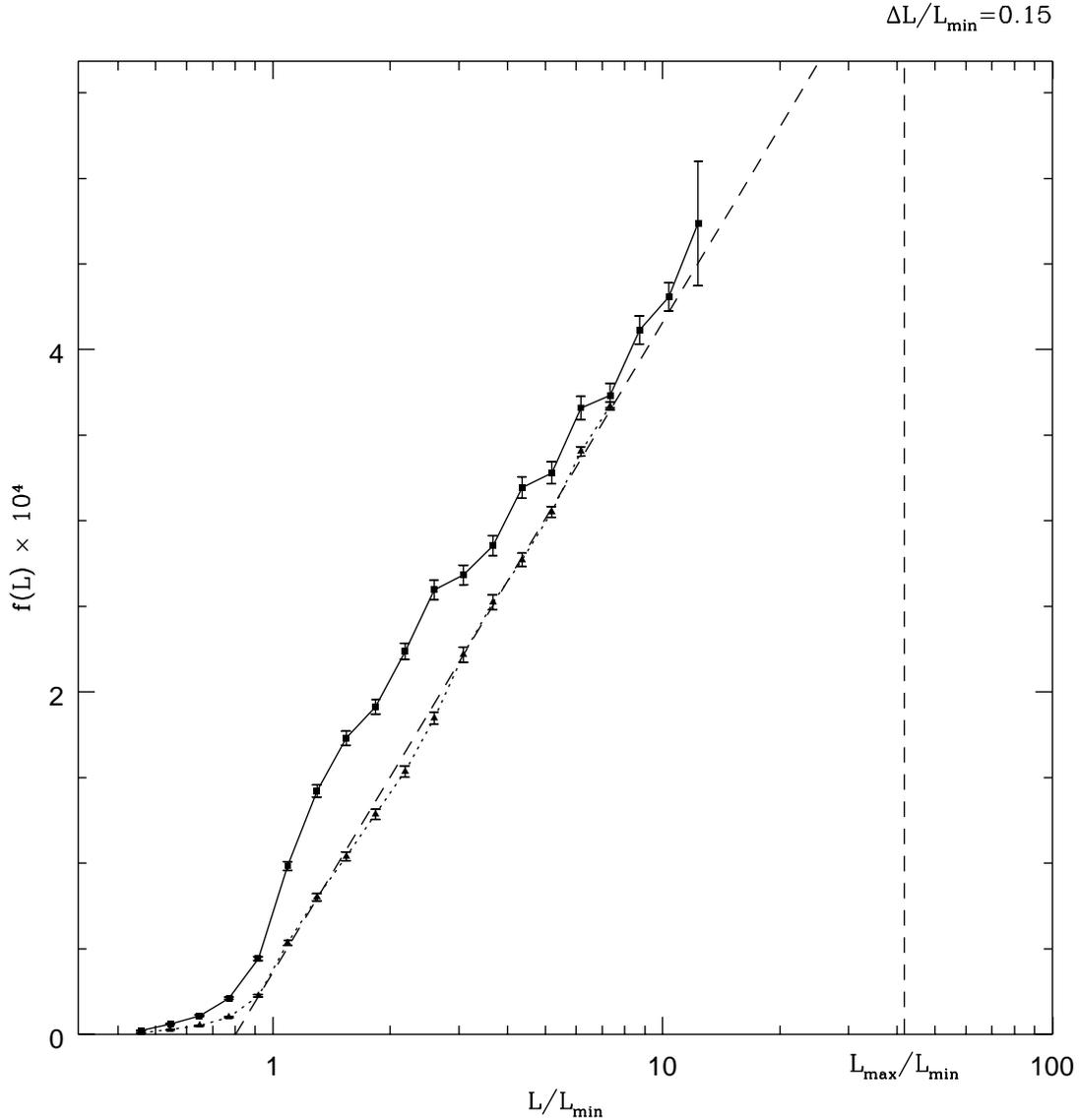}

\caption{
A comparison of resonant (solid curve with square points) 
and non-resonant (dotted curve with triangular points) distribution functions
near $\Lmax$ (marked). The dashed line through the non-resonant $f(L)$ is a
linear (in semi-log coordinates) least-squares fit to the data.
For $L\gg \dLprec\sim 2$,
the two curves appear to merge; only nearer the loss cone is the resonant
enhancement to $f(L)$ significant.
\label{fig_fLcmp}}
\end{figure}

Figure~\ref{fig_fLcmp} compares the resonant (solid)
and non-resonant (dotted) $f(L)$ for a run with $L_0\sim\Lmax$;
a linear least-squares fit to the latter (dashed line) is also plotted.
The figure suggests that for
large $L$ the two curves are essentially equal---the resonant enhancement
becoming significant only near the loss cone. This behavior can be
understood qualitatively as follows. Recalling that even resonant relaxation 
proceeds as a random walk on timescales longer than the orbital precession time
(\S~\ref{sec_resr}), one would expect relaxation over distances much larger
than
$\dLprec$---the effective mean free path in $L$-space---to proceed at almost
identical rates in both resonant and non-resonant systems; their
distribution functions should therefore be nearly coincident in this region.
Near the loss cone, on the other hand (i.e., for $L\lesssim \dLprec$),
the effective diffusion in the
resonant cluster is much higher; here, therefore, the 
resonant $f(L)$ will be flatter than its
non-resonant counterpart, and hence will increasingly dominate over the
latter as the loss cone is approached. Inside the loss cone both curves 
fall exponentially with a scale length of $\dLorb$.
This line of reasoning explains
the gross features visible in Figure~\ref{fig_fLcmp};
in particular, note the apparent turnover in $f_{\rm res}$ for $L\gtrsim
3\Lmin$ (this particular simulation had $\dLprec\sim 2\Lmin$).

\begin{deluxetable}{ccccccccc}
\tablecaption{Resonant and Non-Resonant Tidal Disruption Rates\label{tab_lamb}}

\tablehead{
\colhead{$\cE/\Ecrit$} &
\colhead{$L_{0,{\rm nr}}$\tablenotemark{a}} &
\colhead{$L_{0,{\rm res}}$\tablenotemark{a}} &
\colhead{$\Lmax$\tablenotemark{a}} & \colhead{$\dLorb$\tablenotemark{a}} &
\colhead{$\dLprec$\tablenotemark{a}} &
\colhead{$\lambda_0\times 10^7$} & \colhead{$\lnr\times 10^7$} &
\colhead{$\lres\times 10^7$}
}

\startdata
  8& 2.8& 8& 125& 0.022& $\sim 3$& 0.06& 0.18(0.04)& 5.5(1.3)\nl

  3& 8& 23& 205& 0.14& $\sim 6$& 0.8& 1.4(0.2)& 9(1.5)\nl

  1& 23& 32& 354& 0.80& $\sim 12$& $\sim 25$& 18(2)& 35(6)\nl

  0.3& 64& 91& 614& 5.5& $\sim 24$& 26& 26(1.1)& 26(1.8)\nl
\enddata

\tablenotetext{a}{ In units of $\Lmin$.}
\end{deluxetable}

\subsection{Relativistic Simulations}
\label{sec_gr}

We examined the importance of relativistic precession through two series
of simulations. In each series, the value of $\dLorb/\Lmin$ was held fixed 
while the black hole mass was varied between relatively small values, for which
relativistic effects were negligible, and larger masses for which they
dominated the precession.
The relativistic simulations were performed in a conceptually
simple manner---by merely replacing the Kepler stepper used in the
usual MVS scheme (\S~\ref{sec_tstar}) with one integrating motion in the
Schwarzschild (non-rotating black hole) spacetime.
This is equivalent to including the relativistic corrections to the
Newtonian black hole (i.e., point mass) potential in the ``unperturbed
motion'' part of the symplectic integrator instead of in the perturbation
potential. Although straightforward,
this approach does not treat time dilation effects completely
accurately---time is still considered an absolute, universal quantity. 
Since such effects are noticeable only during the extremely brief passage
through periapse, however, they can be safely neglected in our case.
(In fact, our method is probably overkill; artificially
adding precession at the relativistic rate would likely have been
sufficient.) Because of
this comingling of formalisms, we refer to this scheme as ``semi-relativistic''
symplectic integration; for details on implementation of the Schwarzschild
stepper (which incorporates a high-order approximation to geodesic motion in
the Schwarzschild metric) see Appendix~\ref{app_grstep}.

Figure~\ref{fig_mgr} displays the disruption rate $\lambda$ as a function of
black hole mass for test stars undergoing strongly resonant relaxation 
($\dLorb/\Lmin=0.025$) in the non-relativistic case.
The points $\lambda(M\to 0)$ 
and $\lambda(M\to\infty)$ correspond to purely Newtonian resonant and
non-resonant simulations, respectively. The figure shows very clearly the
quenching effect of relativistic precession at large hole masses: whereas the
scale-free Newtonian simulations predict a factor of $\approx 30$ enhancement in
the tidal disruption rate due to resonant effects (cf. Table~\ref{tab_lamb}),
for a physical hole mass of (say)
$10^6\Msun$, the relativistic simulations show only a factor of
$\approx 5$ net increase.
The ``quenching mass'' at this energy is thus approximately $10^6\Msun$.
Near the critical energy, on the other hand,
relativistic effects from a $10^6\Msun$ black hole would not noticeably wash
out the (much smaller, but more robust)
resonant enhancement to $\lambda$; accomplishing this
requires a much shorter precession period---and hence a larger black
hole.

The second set of relativistic simulations were designed to estimate the
hole mass above which all resonant tidal effects disappear.
Three distinct groups of runs were performed, all with a fixed value of
$\dLorb/\Lmin=0.3$: (1) non-relativistic, non-resonant; (2) non-relativistic,
resonant; and (3) relativistic, resonant. We will define the quenching mass
as the black hole mass for which half the total resonant enhancement 
to the overall tidal disruption is relativistically suppressed.
In this case, the estimate is made more
difficult by the relatively small resonant excess near the critical energy
as well as the large array of runs required.
From the results of $\approx 20$ runs spanning hole masses of
$10^7$ to $10^8$, our best estimate for the quenching mass $M_{\rm Q}$ is
\begin{equation}
M_{\rm Q}=(8\pm 3)\times 10^7\Msun.
\end{equation}
This is also close to the rough estimate of $4\times 10^7\Msun$ given in RT96.

\begin{figure}
\plotone{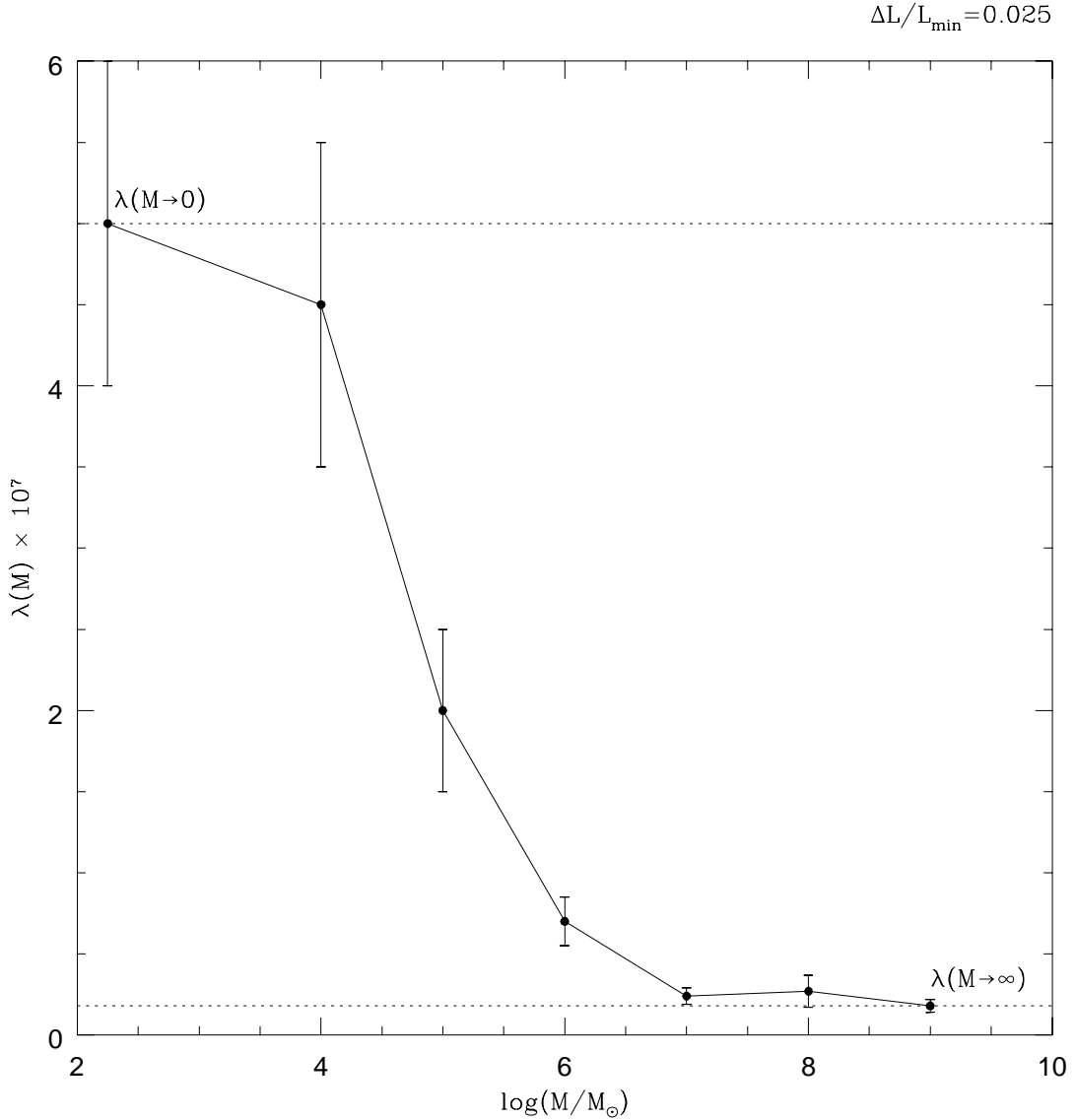}

\caption{
A plot of the fractional disruption rate per orbital period, $\lambda$,
as a function of the black hole mass, $M$, for a series of otherwise identical
resonant clusters.
For very massive holes, the resonant enhancement to $\lambda$ is quenched---the
result of an increasing amount of relativistic precession near
the loss cone as the tidal radius moves closer to the horizon.
When relativistic precession is negligible (leftmost point),
the disruption rate is $\sim 30$ times
higher than in the corresponding non-resonant cluster (rightmost point).
In this particular case, half of this increase is gone by $M\sim 10^6\Msun$;
near the critical energy (where $\dLorb/\Lmin\sim 1$) the quenching mass 
is larger, $\sim 8\times 10^7\Msun$ (see \S~\ref{sec_gr}).
\label{fig_mgr}}
\end{figure}

\section{Discussion}
\label{sec_discuss}

We have shown that resonant tidal disruption strongly enhances the disruption
rate in the empty loss cone limit, by a factor of $\approx \Lmin/\Delta L$; the
total disruption rate for the entire cluster is approximately double the
non-resonant rate.  We also found that for hole masses
$M\gtrsim 8\times 10^7\Msun$, the latter increase is eliminated by 
relativistic precession; sufficiently far into the diffusion limit, this
quenching effect is also present for smaller black holes.
These results quantitatively confirm the order-of-magnitude estimates given in
RT96 and highlight the potential importance of resonant relaxation in
determining the structure of galactic nuclei at small radii. 

Other processes capable of modifying loss cone dynamics include black hole
wander and tidal capture.  Black hole wander (\cite{lint80}; \cite{you77};
\cite{bahw76})---the result of two-body relaxation acting on the
hole itself---is a Brownian motion-like process which causes the hole to
undergo minor excursions from the true dynamical center of the nucleus.
Like resonant tidal disruption, BH wander is never important in the
full loss cone regime---not even the existence of a loss cone, much less its
precise location, is of consequence in this case. In the empty loss cone case,
BH wander can modestly enhance the local non-resonant disruption rate by a 
logarithmic factor (\cite{you77}); however, we do not expect resonant
relaxation and BH wander to significantly reinforce each other in this regard.
This is because resonant nuclei have relatively low densities and hence large
$\rcrit$; the low velocity dispersions here then imply loss cone impact
parameters $b_{\rm t}\sim \Lmin/v\gg r_{\rm w}$, where $r_{\rm w}$ is the rms
BH wandering radius (even in the relatively dense Galactic Center, for example,
$b_{\rm t}\sim 3 r_{\rm w}$ near $\rcrit$). At smaller radii, by contrast,
the tightly bound cusp stars have orbital periods short compared to 
the wandering timescale, and therefore follow the hole adiabatically. There
could be an intermediate region in which BH wander and resonant effects
reinforce each other, but in general this does not happen.

Tidal capture, on the other hand, will further enhance resonant disruption
rates just as it enhances non-resonant tidal disruption. Tidal capture
(\cite{dieet95}; \cite{novpp92}; \cite{ree88}; \cite{frar76}) occurs when a
star passes close to, but outside of, its tidal radius; although it escapes
immediate destruction, the tides raised on the star dissipate enough
orbital kinetic energy inside the star
to substantially reduce the semi-major axis of the
orbit (the pericentric distance remains nearly fixed).
This process continues until either the orbit circularizes or
the star disrupts; in the case of a massive black hole, the studies indicate
that the latter will occur.
(We remark that the consequence of tidal capture in $\cE$-space is
conceptually similar to that of resonant relaxation in $L$-space: a systematic
drift that overwhelms the diffusion induced by two-body effects.)
Thus tidal capture effectively increases the value
of $\Lmin$ (by a factor of $\approx 1.5$; Novikov et al.\ 1992);
the result is a further doubling of the stellar disruption rate. Since this
gas is, on average, tightly bound to the hole,
the increase in accretable gas will be even
greater. This further improves the ability of the loss cone
to power at least low-level activity in galactic nuclei.

How important a role will resonant tidal disruption play in normal galactic
nuclei? Table~\ref{tab_lcprop} lists the nuclear properties of several black
hole candidates, including estimates of the critical radius, $\rcrit$,
and the cluster mass fraction at that location, $\mucrit$.
The table suggests that typical galactic nuclei are
non-resonant at their critical radii (recall that resonant
effects dominate only for mass fractions $\mu\lesssim 0.1$); resonant
enhancement to the {\it overall} disruption rates in these nuclei will
therefore be quite small. At radii a few times smaller than $\rcrit$, however, 
most galaxies in the table {\it are} dominated by resonant relaxation, and
these regions can be resolved in several instances. Because erosion of the
density cusp inside $\rcrit$ progresses so slowly, however
(see \S~\ref{sec_lcone}), its detection is unlikely even after accounting for
resonant effects. We therefore conclude that resonant tidal disruption 
will not visibly alter nuclear density profiles at currently accessible
angular resolutions.

Another astrophysical system in which resonant loss cone effects
are important is
the Oort cloud. Although several processes contribute, the dynamical evolution
of Oort cloud comets in the outer Solar System ($r\sim 10^4$ AU) is
controlled mainly by the Galactic tidal field (\cite{wiet97}; \cite{heit86}).
The tidal perturbation to the nearly-Keplerian
potential of the Solar System resonantly relaxes the angular momentum of the
comet orbits, occasionally removing so much of it that the comets plunge into
the inner Solar System, where they may become visible as long period comets.
The loss cone in this situation arises from the ability of the giant planets
(particularly Jupiter and Saturn) to scatter incoming comets onto either
hyperbolic or relatively tightly bound orbits---in either case causing the
comets to be ``lost'' from the Oort cloud.
As for nuclear clusters around MBHs, the rate at which comets reach the
scattering loss cone---as well as the equilibrium Oort cloud
distribution function at small $L$ (which directly determines the 
frequency of `new' long period comets)---is set by resonant loss
cone dynamics.

The relative difficulty of the numerical calculations highlights the 
generic advantage of using a continuum approach in this class of problems,
and indicates that the development of a resonant Fokker-Planck formalism
certainly merits further attention. The inherently non-linear nature of
resonant relaxation and the associated complexity of its analytic description,
however, suggests that it is likely to be a highly non-trivial undertaking!
Further exploration of this issue is beyond the scope of this paper.

\begin{deluxetable}{cccccccc}
\tablecaption{Loss Cone Properties of Nearby Black Hole Candidates
  \label{tab_lcprop}}

\tablehead{
\colhead{Galaxy} & \colhead{$D$ (Mpc)} & \colhead{$M_8$} &
\colhead{$\rho_6$\tablenotemark{a}} &
\colhead{$n$\tablenotemark{a}} & \colhead{$\rcrit$ ($''$)} & 
\colhead{$\rcrit$ (pc)} & \colhead{$\mucrit$}
}

\startdata
  Milky Way& 0.01& 0.025& 1& 1.8& 7& 0.3& 0.3\nl
  M32& 0.7& 0.02& 0.2& 2& 0.1& 0.4& 0.3\nl
  M31& 0.7& 0.3& 0.1& 1& 1.5& 5& 0.4\nl
  NGC 3377& 9.9& 0.8& 0.2& 1.3& 0.25& 12& 2\nl
  NGC 3115& 8.4& 20& 0.6& 1.75& 8& 350& 4\nl
  M87& 15.3& 30& 0.01& 1.2& 8& 650& 4\nl
\enddata

\tablenotetext{a}{ $\rhos(r)=10^6\rho_6(r/1\psc)^{-n}\Msun\psc^{-3}$}
\end{deluxetable}

\acknowledgements

We thank Scott Tremaine for helpful discussions and a careful reading of the
manuscript.
This work was supported by NSERC and a Jeffrey L. Bishop Fellowship to K.~R.

\appendix
\section{Semi-Relativistic Symplectic Integration}
\label{app_grstep}

As discussed in \S~\ref{sec_gr}, some simulations incorporated relativistic
corrections to the test stars' 
orbital motion to determine the hole masses for which 
resonant tidal disruption is shut off by relativistic precession.
Operationally this was accomplished by replacing the Kepler
stepper normally used in the MVS integration scheme,
which solves for motion in the Newtonian one-body problem,
with a Schwarzschild stepper approximating the motion around a Schwarzschild
(i.e., non-rotating) black hole. In this appendix
we describe the construction of the Schwarzschild stepper itself.

In the Newtonian case all five orbital elements---the semi-major axis, $a$,
eccentricity, $e$, inclination, $i$, argument of periapse, $\omega$, and
argument of the ascending node, $\Upsilon$---are constants of the motion; hence
orbits are closed and do not precess in the absence of perturbations.
In the Schwarzschild case, continued spherical symmetry
(which ensures that the orbital motion is planar)
combined with energy and (scalar) angular momentum conservation imply
that four of these elements---$a$, $e$, $i$, and $\Upsilon$---remain
constant and
well-defined even for highly relativistic orbits; $\omega$, however, is {\it
not} fixed but precesses by an amount
$\Delta\omega \approx 6\pi/[a(1-e^2)]$ rad per orbit in the weak field limit.
(The exact expression is
$\Delta\omega=2[\beta \ellK(\beta^2 \delta e)-\pi]$, where
$\delta=1/[a(1-e^2)]$, $\beta=2/[1-2\delta(3-e)]^{1/2}$, and $\ellK(m)$ is the
complete elliptic integral of the first kind.)

A Kepler stepper advances the orbital position $\mbf{x}$ and velocity $\mbf{v}$
by a time $\Delta t$ through solution of
some form of Kepler's Equation, which determines the
orbital phase as a function of time.
The standard form of Kepler's Equation for elliptical orbits is
\begin{equation}
M=E-e\sin E,
\end{equation}
where $M=2\pi(t-\tp)/\torb$ is the mean anomaly, $E$ is the eccentric anomaly
(related to the radius by $r=a(1-e\cos E)$), $\tp$ is a time of pericentric
passage, and $\torb=2\pi a^{3/2}$ is the orbital period.
Conceptually, a relativistic geodesic stepper advances the
4-position $\mbf{X}$ and 4-momentum $\mbf{P}$ over some interval of
coordinate time $\Delta t$; however, since the Schwarzschild metric is static
it is sufficient in this case to use just the 3 spatial components of
position and momentum (or velocity) to advance orbits, even ones passing
near the horizon.  One practical
disadvantage of using the true relativistic 3-velocity is that
calculation of the orbital elements requires setting up and solving a cubic
polynomial, which is time-consuming (as it must be redone every time step).
For this reason we created a `semi-relativistic' Schwarzschild stepper which
returns the true orbital 3-position but
an effective 3-velocity for which application of the simpler Newtonian formulae
yield the correct (relativistic) orbital elements; in the weak field limit, 
both of these
vectors become identical to their non-relativistic counterparts
$\mbf{x}$ and $\mbf{v}$. Our semi-relativistic stepper can thus be used as
an exact replacement for the Kepler stepper in the integration of
moderately-relativistic orbits.

Implementing a Schwarzschild stepper requires an analog of Kepler's Equation
which is valid for relativistic orbits; generalized expressions for $\torb$
and the true anomaly, $\nu$ (defined as the physical angle traversed in the
orbital plane since periapse), are
also needed. Rauch (1997) has derived these formulae for orbits in the Kerr
(rotating black hole) spacetime accurate to second order in $\delta$
(see above), assuming $\delta\ll 1$.
For Schwarzschild the results (with errors $O(\delta^3)$) simplify to
\begin{eqnarray}
\torb& =&
  2\pi a^{3/2}\left\{1+\left(\frac{3}{a}\right)\left[1+\left(2+\frac{5}{2}
  \sqrt{1-e^2}\right)\delta\right]\right\}\, ,\\
\nu(\chi)& =&
  \left\{1+3\delta+\left[54+3e^2\right]\frac{\delta^2}{4}\right\}\chi+
  \left\{1+\left[9+\frac{3}{4}e\cos\chi\right]\delta\right\}\delta e\sin\chi\, ,
\end{eqnarray}
and $M=E-{\tilde e}\sin E+\Delta E$, where
$\Delta E=\frac{15}{2}(\chi-E)(1-e^2)^{3/2}\delta^2$ and
\begin{equation}
{\tilde
e}=e\left\{1-\left(\frac{3}{a}\right)\left[1+
  \left(-10+18e^2+15\sqrt{1-e^2}\right)\frac{\delta}{6}
  \right]\right\}\, .
\end{equation}
Here $\chi$ is the ``relativistic anomaly,'' defined according to
$r=a(1-e^2)/(1+e\cos\chi)$---hence $\chi$ advances by exactly $2\pi$ from one
periapse to the next, regardless of the amount of orbital precession (which is
$\Delta\omega=\nu(2\pi)-2\pi\approx 6\pi\delta$, in agreement with the value
given above); for strictly Keplerian orbits (i.e., $\delta\to 0$), $\chi=\nu$.
Note that since $0\le {\tilde e}< e < 1$, the modified Kepler's
Equation can be solved economically by finding the solution for $\Delta E\to 0$
and then correcting for the finite value of $\Delta E$. We found the
resulting stepper to be only a factor of two slower than its Newtonian
counterpart and to give excellent results (compared to accurate
integrations of the Schwarzschild problem) for radii $r\gtrsim 10$, at which
point the systematic error in the precession rate was $\approx 25\%$.

\section{Chaos in MVS Integration of Nearly-Radial Orbits}
\label{app_chaos}

By construction the MVS integration scheme (see \S~\ref{sec_tstar})
is exact in the limit of unperturbed Keplerian motion.
As a symplectic mapping, the method might also be expected to be
free of systematic, long-term growth of energy error in the slightly perturbed
problem (i.e., for nearly-Keplerian motion)---this property being one of the
key advantages of such algorithms. Applied to our simulations, however, the
MVS integrator was found to exhibit unbounded error growth even for very small
perturbations, in some cases causing test star orbits to become unbound in as
few as 100 orbital periods (the physical energy relaxation timescale being
orders of magnitude longer). As we explain below, the instability is related 
to the very high eccentricities of the test star orbits; qualitatively
similar integrations (e.g., tracing the orbital evolution of Oort cloud comets)
are therefore likely to suffer from the same problem.
In this section we briefly discuss the problem, and propose a modified
integration scheme which remains robust under such circumstances.

Our integrations are somewhat unique for two reasons: the test star orbits
are all highly eccentric (generally, $e\gtrsim 0.999$), and the perturbations 
from the background cluster are very nearly constant over scales of order the
pericentric distance---the perturbations are 
spatially well-resolved with a stepsize $\Delta t/\torb\sim 10^{-3}$, taking
the form of a small, fixed tidal force near periapse.
Hence there is no {\it a priori} 
reason to integrate using a stepsize small enough to resolve periapse
($\Delta t/\torb\sim (1-e)^{-3/2}\sim 10^{-5}$), which
would be impractical in any case---recall that symplectic integrators do not
easily admit variable stepsizes; this is (in principle!)
the great advantage of the MVS
method for nearly-Keplerian problems. For stepsizes not resolving periapse,
however, it was found that significant jumps in the energy, centered around
periapse, occurred each orbit. This
produced a linearly growing energy error---even
in simulations where the background masses were held fixed, for which
energy should be strictly conserved. In our case, it was possible to work
around the problem by artificially truncating the cluster at a relatively
large radius--- $\sim a/30$ for test stars with semi-major axis $a$---for
which the residual
tidal force at periapse was small enough that physical energy relaxation
dominated that spuriously produced by the MVS method; angular momentum
relaxation is nearly unchanged as it is controlled by the largest-scale
fluctuations in the background cluster. Although this proved adequate,
it would be more elegant (and generically more useful) 
to modify the MVS scheme so that it did not suffer from this problem; 
although a detailed treatment is beyond the scope of this paper, we will
mention one possible solution.

Since the perturbing force is almost constant near periapse---where effectively
all of the erroneous relaxation occurred---an excellent model
analogy to our situation is provided by the Stark
problem, in which the perturbing force is strictly constant in magnitude and
direction over the entire orbit.
Besides its simplicity, the Stark problem has
the added advantages of being integrable and having a conserved energy; this
guarantees that any energy relaxation or chaotic behavior found in numerical
integrations of the problem are an artifact of the integrator itself.
Sample integrations using the
MVS scheme confirmed that for stepsizes resolving periapse, there was no
systematic growth in the energy error, as nominally expected from a symplectic
scheme; unresolved orbits, on the other hand, generally had erratic,
unbounded growth of energy error, even for extremely small perturbations.
More direct (though informal) 
evidence for dynamic chaos in the integrator was found in the
sensitivity of the relaxation curves to the precise initial conditions used,
which was not present in the stable integrations.

The close correspondence between the Stark problem and our simulations suggests
that a more robust integrator might result
by replacing the Kepler stepper used
in the MVS method with a Stark stepper solving that problem exactly (which
can be done in parabolic coordinates using elliptic functions and integrals).
In contrast to the standard MVS scheme, which is exact in the limit of zero
perturbations, the Stark integrator is exact in the limit of a constant
(possibly large) perturbation force. A brief investigation revealed that
the new method, as expected,
does not suffer from the aberrant relaxation present in MVS scheme---
even for large stepsizes that do not resolve the passage through periapse.
On the other hand, the practicality of a Stark integrator depends
on the speed of the Stark stepper itself, which will be much slower
(and more tedious to implement) than the Kepler stepper it replaces.
Based on the encouraging initial tests,
an efficient implementation of the proposed Stark integrator is currently
being undertaken; detailed results will be reported elsewhere.


\begin{thebibliography}{}

\bibitem[Bade, Komossa, \& Dahlem 1996]{badkd96}
Bade N., Komossa S., Dahlem M., 1996, \aap, 309, L35

\bibitem[Bahcall \& Wolf 1976]{bahw76}
Bahcall J. N., Wolf R. A., 1976, \apj, 209, 214

\bibitem[Carter 1992]{car92}
Carter B., 1992, \apj, 391, L67

\bibitem[Carter \& Luminet 1983]{carl83}
Carter B., Luminet J.-P., 1983, \aap, 121, 97

\bibitem[Cohn \& Kulsrud 1978]{cohk78}
Cohn H., Kulsrud R. M., 1978, \apj, 226, 1087

\bibitem[Danby 1992]{dan92}
Danby J. M. A., 1992, Fundamentals of Celestial Mechanics. Willmann-Bell Inc.,
Richmond

\bibitem[Diener et al.\ 1995]{dieet95}
Diener P., Kosovichev A. G., Kotok E. V., Novikov I. D., Pethick C. J., 1995,
\mnras, 275, 498

\bibitem[Duncan \& Shapiro 1983]{duns83}
Duncan M. J., Shapiro S. L., 1983, \apj, 268, 565

\bibitem[Eckart \& Genzel 1996]{eckg96}
Eckart A., Genzel R., 1996, \nat, 383, 415

\bibitem[Eracleous, Livio, \& Binette 1995]{eralb95}
Eracleous M., Livio M., Binette L., 1995, \apj, 445, L1

\bibitem[Eracleous et al.\ 1995]{eraet95}
Eracleous M., Livio M., Halpern J., Storchi-Bergmann T., 1995, \apj, 438, 610

\bibitem[Evans \& Kochanek 1989]{evak89}
Evans C. R., Kochanek C. S., 1989, \apj, 346, L13

\bibitem[Frank 1978]{fra78}
Frank J., 1978, \mnras, 184, 87

\bibitem[Frank \& Rees 1976]{frar76}
Frank J., Rees M., 1976, \mnras, 176, 633

\bibitem[Grupe et al.\ 1995]{gruet95}
Grupe D., Beuermann K., Mannheim K., Bade N., Thomas H.-C., de Martino D.,
  Schwope A., 1995, \aap, 299, L5

\bibitem[Heisler \& Tremaine 1986]{heit86}
Heisler J., Tremaine S., 1986, Icarus, 65, 13

\bibitem[Hills 1975]{hil75}
Hills J. G., 1975, \nat, 254, 295

\bibitem[Hills 1988]{hil88}
Hills J. G., 1988, \nat, 331, 687

\bibitem[Hills 1991]{hil91}
Hills J. G., 1991, \aj, 102, 704

\bibitem[Khokhlov \& Melia 1996]{khom96}
Khokhlov A., Melia F., 1996, \apj, 457, L61

\bibitem[Kinoshita, Yoshida, \& Nakai 1991]{kinyn91}
Kinoshita H., Yoshida H., Nakai H., 1991, Cel. Mech., 50, 59

\bibitem[Kochanek 1994]{koc94}
Kochanek C. S., 1994, \apj, 422, 508

\bibitem[Kormendy \& Richstone 1995]{korr95}
Kormendy J., Richstone D., 1995, \araa, 33, 581

\bibitem[Kormendy et al.\ 1996]{koret96}
Kormendy J. et al., 1996, \apj, 459, L57

\bibitem[Laguna et al.\ 1993]{laget93}
Laguna P., Miller W. A., Zurek W. H., Davies M. B., 1993, \apj, 410, L83

\bibitem[Lauer et al.\ 1995]{lauet95}
Lauer T. R. et al., 1996, \aj, 110, 2622

\bibitem[Lightman \& Shapiro 1977]{ligs77}
Lightman A. P., Shapiro S. L., 1977, \apj, 211, 244

\bibitem[Lin \& Tremaine 1980]{lint80}
Lin D. N. C., Tremaine S., 1980, \apj, 242, 789

\bibitem[Luminet \& Barbuy 1990]{lumb90}
Luminet J.-P., Barbuy B., 1990, \aj, 99, 838

\bibitem[Lynden-Bell 1969]{lyn69}
Lynden-Bell D., 1969, \nat, 223, 690

\bibitem[McMillan, Lightman, \& Cohn 1981]{mcmlc81}
McMillan S. L. W., Lightman A. P., Cohn H., 1981, \apj, 251, 436

\bibitem[Novikov, Pethick, \& Polnarev 1992]{novpp92}
Novikov I. D., Pethick C. J., Polnarev A. G., 1992, \mnras, 255, 276

\bibitem[Quinlan \& Hernquist 1997]{quih97}
Quinlan G. D., Hernquist L., 1997, preprint

\bibitem[Quinlan, Hernquist, \& Sigurdsson 1995]{quihs95}
Quinlan G. D., Hernquist L., Sigurdsson S., 1995, \apj, 440, 554

\bibitem[Rauch 1997]{rau97}
Rauch K. P., 1997, \apj, submitted

\bibitem[Rauch \& Tremaine 1996]{raut96}
Rauch K. P., Tremaine S., 1996, NewA, 1, 149

\bibitem[Rees 1988]{ree88}
Rees M., 1988, \nat, 333, 523

\bibitem[Roos 1992]{roo92}
Roos N., 1992, \apj, 385, 108

\bibitem[Shapiro \& Marchant 1978]{sham78}
Shapiro S. L., Marchant A. B., 1978, \apj, 225, 603

\bibitem[Spitzer \& Hart 1971]{spih71}
Spitzer L., Hart M. H., 1971, \apj, 164, 399

\bibitem[Stoeger, Pacholczyk, \& Stepinski 1992]{stops92}
Stoeger W. R., Pacholczyk A. G., Stepinski T. F., 1992, \apj, 391, 550

\bibitem[Storchi-Bergmann et al.\ 1995]{stoet95}
Storchi-Bergmann T., Eracleous M., Livio M., Wilson A., Filippenko A.,
  Halpern J., 1995, \apj, 443, 617

\bibitem[van der Marel et al.\ 1997]{vdmet97}
van der Marel R. P., de Zeeuw P. T., Rix H.-W., Quinlan G. D., 1997,
\nat, 385, 610

\bibitem[Wiegert \& Tremaine 1997]{wiet97}
Wiegert P., Tremaine S., 1997, Icarus, submitted

\bibitem[Wisdom \& Holman 1991]{wish91}
Wisdom J., Holman M., 1991, \aj, 102, 1528

\bibitem[Young 1977]{you77}
Young P. J., 1977, \apj, 215, 36

\end{thebibliography}
\end{document}